\documentstyle[11pt,newpasp,twoside,epsf]{article}
\def\edcomment#1{\iffalse\marginpar{\raggedright\sl#1\/}\else\relax\fi}
\reversemarginpar

\begin{document}
\title{Warm Ionized Gas on the Outskirts of Active and Star-Forming Galaxies}
\author{Sylvain Veilleux}
\affil{Department of Astronomy, University of Maryland, College Park, MD 20742}

\begin{abstract}
The preliminary results from a deep emission-line search for warm
ionized material in the halos of nearby active and star-forming
galaxies are presented.  The origin of this gas is discussed in the
context of galaxy formation and evolution.
\end{abstract}

\section{Introduction}

The need for a comprehensive survey of the warm ionized medium in
the local universe can hardly be overstated. This gas phase may
contribute significantly to the local baryon budget (e.g., Fukugita,
Hogan, \& Peebles 1998). While an important fraction of this material
is almost certainly in the form of intergalactic clouds not related to
any individual galaxy, some may inhabit the dark matter halos of
galaxies.

This gas phase is a key witness of galaxy formation and evolution. In
a hierarchical CDM universe (e.g., Jenkins et al.  1998) most of the
activity associated with galaxy formation takes place at $z$ $\ga$ 1,
except perhaps in the outer reaches of galaxies where ``primordial''
gas may still be accreting today. Some of this material will
necessarily be ionized by the metagalactic ionizing radiation, and
possibly by local sources of ionizing radiation such as active
galactic nuclei (AGN) and starburst galaxies. In this picture, the
warm ionized gas on the outskirts of galaxies represents left-over
debris associated galaxy formation.

The warm ionized gas is an excellent probe of the feedback processes
taking place in galaxies. Ionizing radiation and energy injection from
star-forming regions and quasars may severely limit the amount of star
formation and affect galaxy evolution. Galactic winds may blow out
through the gaseous halos of galaxies and into the IGM, enriching and
heating the intergalactic environment in the process. These winds may
be responsible for the well-known mass-metallicity relation in
galaxies (e.g., Larson \& Dinerstein 1975; Vader 1986; Franx \&
Illingworth 1990). The warm ionized gas phase in these winds is
primarily made of the cool/warm ISM originally belonging to the host
galaxies.

This paper describes the preliminary results of a deep emission-line
survey of the local universe. The main goals of this survey are to
search for warm (T $\approx$ 10$^4$ K) ionized gas on the outskirts of
nearby galaxies, and study the properties of this gas to find out its
origin and overall importance. For lack of space, the present
discussion focusses only on the warm ionized halos of active and
star-forming galaxies; we do not discuss our recent results on the
warm ionized edges of disk galaxies and on isolated extragalactic H~I
clouds (this last topic is discussed by S. Vogel in this volume; see
also Weymann et al. 2001).

\section{Observations}

Over the past few years, our group has obtained deep emission-line
images of several starburst and active galaxies using the Taurus
Tunable Filter (TTF)\footnote{http://www.aao.gov.au/ttf} on the 3.9m
Anglo-Australian and 4.2m William Herschel Telescopes.  This
instrument is uniquely suited for this type of survey, combining wide
field of view with outstanding narrow-band imaging capabilities over a
broad range in wavelength (3500 \AA\ -- 1.0 $\mu$m) and bandpass (10
-- 100 \AA). The TTF uses high-performance, low-order Fabry-Perot
etalons equipped with long-range piezo-electric transducers now
sufficiently reliable to allow the parallel plates to be scanned over
a physical spacing of 4 $\mu$m (5 wavelengths or 10 interference
orders in the I band) and to allow parallelism to be maintained down to
spacings of $\sim$ 1 $\mu$m. To maximize sensitivity to faint flux
levels, we are using the TTF in the so-called ``charge
shuffling/frequency switching'' mode.  The basic idea is to move
charge up and down within the detector at the same time as switching
between two discrete frequencies with the tunable filter.  The chip is
read out only once.  As a result, the TTF can produce {\em
simultaneous} continuum and emission-line images that reach flux
levels of order a few x 10$^{-18}$ erg s$^{-1}$ cm$^{-2}$
arcsec$^{-2}$ ($\varepsilon_m \sim$ 1 cm$^{-6}$~pc), an order of
magnitude fainter than typical narrow-band images published on normal
or active galaxies. Many of the images were obtained in a straddle
mode, where the off-band image is made up of a pair of images that
``straddle'' the on-band image in wavelength (e.g., $\lambda_1$ = 6500
\AA\ and $\lambda_2$ = 6625 \AA\ for rest-frame H$\alpha$); this
greatly improves the accuracy of the continuum removal since it
corrects for slopes in the continuum, underlying absorption features,
etc.

Complementary long-slit optical spectra were also obtained for some
objects to clarify the origin and source of ionization of the warm
ionized material. When available, multiwavelength and HST data were
used to track the hot (ROSAT, Chandra) and relativistic (VLA) gas
phases and improve on the spatial resolution of our ground-based data.

\section{Results}

\subsection{Normal Star-Forming Disk Galaxies}

This portion of the survey is part of Scott Miller's Ph.D. thesis at
the University of Maryland (Miller 2001). Twenty non-active,
non-interacting spirals were selected for this study based on their
proximity ($z$ $<$ 0.01; for good spatial resolution: $<$ 200 pc
arcsec$^{-1}$), angular size ($<$ 10' = field of view of the TTF), and
inclination ($i$ $>$ 75$^\circ$).

Thick ionized disks and/or filamentary structures are detected out to
a few kpc from the planes of several galaxies in our sample.  This is
consistent with the results from earlier imaging studies (e.g.,
Rossa \& Dettmar 2000; 
Collins \& Rand 2001, and references therein). Both the mass
and extent of the extraplanar material in these galaxies appear to be
correlated with the {\em local} surface density of star formation
activity in the disk.

Multi-line imaging with the TTF (see Fig. 1 for one example) indicates
that the emission-line ratios of the high-$\vert z \vert$ ionized gas
differ considerably from those of H~II regions. For instance, the
[N~II] $\lambda$6583/H$\alpha$ ratio in the extraplanar gas often
exceeds unity. Our complementary long-slit data confirm these results,
showing a general increase of the [N~II]/H$\alpha$, [S~II]/H$\alpha$,
and [O~I]/H$\alpha$ ratios with increasing heights in most galaxies.
The trends seen in these ratios of collisionally-excited lines to
recombination lines indicate that the amount of heating per ionization
increases with increasing vertical distance from the disk.

\begin{figure}
\vskip 3.5in
\caption{Multi-line imaging of normal edge-on disk galaxies. NGC
891. (left) H$\alpha$. (center) [N~II] $\lambda$6583. (right) [N~II]
$\lambda$6583/H$\alpha$ ratio map.}
\end{figure}
 
We are currently modelling these variations using the photoionization
code CLOUDY. Radiation from OB stars in the disk appears to be capable
of explaining most of these line ratios, as long as one takes into
account the multi-phase nature of the ISM, the possible depletion of
certain gas-phase abundance of metals onto dust grains, and the
absorption and hardening of the stellar radiation field as it
propagates through the dusty disk. However, positive vertical
[O~III]/H$\alpha$ gradients in a few galaxies suggest the presence of
an additional source of ionization in these objects (e.g., shocks,
turbulent mixing layers).

\subsection{Starburst and Active Galaxies}

Over the history of the universe, AGN- and starburst-driven winds may
have had a strong cumulative impact on the thermal and chemical
evolution of galaxies and their environment (see Cecil 2000, Veilleux
2000, Veilleux et al. 2001, and contributions from T. Heckman,
C. Martin, and M. Lehnert to this volume for a summary of the
situation). So far, very little is known on high-$z$ winds (Pettini et
al. 2000 and references therein). Studies of their local counterparts
have the clear advantage of increased spatial resolution and
sensitivity to faint low-surface-brightness features. The recent
detection of H$\alpha$ filaments (the ``cap'') and diffuse soft X-ray
emission out to 11.6 kpc to the north of the prototypical
starburst/superwind galaxy M82 emphasizes the need for surveying large
areas around superwind galaxies to constrain the size, energetics, and
impact of these superwinds. The radial velocity of the cap (50 -- 200
km s$^{-1}$) exceeds the local escape velocity. Thus, the blowout of
the nuclear superwind and the cap of H$\alpha$-emitting plasma that
delineates its most distant optically visible component is likely to
inject metal-enriched matter into the intergalactic medium of the
M81/M82/NGC~3077 system (Devine \& Bally 1999; Lehnert, Heckman, \&
Weaver 1999). The existence of a 30-kpc limb-brightened
``double-bubble'' system in Arp~220 (Heckman et al. 1987, 1996),
powered by a luminous starburst or dust-enshrouded QSO, is another
indication that superwinds often have large ``spheres of influence''.

Arguably the most spectacular example of a superwind (superbubble) in a
nearby galaxy lies in the core of NGC~3079, an otherwise normal
looking edge-on SBc galaxy at 17 Mpc. Detailed Fabry-Perot, HST, and
VLA studies of this galaxy by our group (Veilleux et al. 1994; Cecil
et al. 2001) reveal a system of ionized strands $\sim$ 0$\farcs$3
(25 pc) wide which emerge from the nuclear region as five distinct gas
streams with velocity gradients and dispersions of hundreds of km
s$^{-1}$ (Fig. 2). The pattern of magnetic fields and the gas
kinematics suggest a wind of mechanical luminosity 10$^{43}$ erg
s$^{-1}$ that has stagnated in the galaxy disk at a radius $\sim$ 800
pc, flared to larger radii with increasing height as the balancing ISM
pressure reduces above the disk, and entrained dense clouds into a
vortex. The observed sphere of influence of this wind is at least
$\sim$ 3--5 kpc, based on the detection of X-shaped optical filaments
and X-ray emission extending on that scale. But circumstantial
evidence for a much larger outflow exists in this galaxy. Irwin et
al. (1987) have indeed argued that the outflow extends out to at least
$\sim$ 50 kpc, based on the presence of an elongated HI tail in the
nearby dwarf S0 galaxy NGC 3073 which is remarkably aligned with the
nucleus of NGC~3079. Irwin et al. argue that the ram pressure from a
wind freely expanding ($n \propto r^{-2}$) out of the nucleus of
NGC~3079 would be more than sufficient to produce the H~I tail. The
existence of this purported giant wind has not yet been confirmed by
other means, however.

\begin{figure}
\vskip 2.9in
\caption{H$\alpha$ + [N~II] line emission map of the nuclear
superbubble in NGC~3079 obtained with HST WFPC2 by Cecil et
al. (2001). }
\end{figure}
 
In some AGN and starbursts, the intense nuclear radiation field leaks
out of the center and photoionizes the material on the outskirts of
the host galaxies. The geometry of the nuclear region (e.g., opening
angle of the accretion disk in AGN, shape of the star-forming
molecular disk in nuclear starbursts) and nature of the host galaxy
(cool disk galaxy vs.  hot spheroid-dominated system) determine the
angular dependence of the ionizing radiation field on large scales.
One of the best cases of such ``ionization cone'' is seen in the
Seyfert galaxy NGC~5252.  The line-emitting gas in this galaxy is
distributed in a wide-angle bicone with opening angle of $\sim$
75$^\circ$ that extends over $\sim$ 40 kpc and consists of a complex
network of filamentary strands (Tadhunter \& Tzvetanov 1989; Wilson \&
Tzvetanov 1994). A detailed Fabry-Perot study of this region (Morse et
al.  1998) shows complex kinematics which are best explained as the
superposition of two inclined rotating disk, probably originating from
a past galaxy merger event.

Our TTF survey has revealed new examples of large-scale ionization
structures. Figure 3 shows our recent results on the prototypical
Seyfert 2 galaxy NGC~1068. Here, a line-emitting filamentary complex
is detected at H$\alpha$ out to $\sim$ 12 kpc from the nucleus,
slightly beyond the H~I edge of this galaxy [as measured from the
recent H~I maps of Brinks (2001)]. Multi-line imaging of this galaxy
with the TTF detects this complex in several other lines including
[N~II] $\lambda$6583, [S~II] $\lambda\lambda$6716, 6731, and [O~III]
$\lambda$5007. The biconical geometry of the complex and the relative
strengths of the emission lines suggest that the central AGN is
contributing to the ionization of this material. This bicone is
roughly aligned with the nuclear outflow/jets and ionization cone seen
on smaller scales (e.g., Cecil, Tully, Bland 1990; Macchetto et
al. 1994). It is not clear at present whether the line-emitting
filaments represent disk material that is simply being illuminated by
the AGN or whether the ionized gas is taking part in a dynamical event.

\begin{figure}
\vskip 6in
\caption{ NGC~1068 in H$\alpha$. H$\alpha$ emission is detected for
the first time out to the H~I edge of this galaxy. The emission is
more visible in the NE (upper left) quadrant; fainter emission is also
detected SW of the nucleus. The field of view of this image is about
6$\farcm$3 or 26 kpc at the distance of NGC~1068. (Shopbell, Bland-Hawthorn,
\& Veilleux 2001, in prep.).}
\end{figure}
 
Figure 4 shows a deep H$\alpha$ + [N~II] $\lambda$6583 + continuum
image of the Seyfert 1 galaxy NGC~7213 obtained with the TTF on the
AAT.  This image reveals the presence of a line-emitting filament
located $\sim$ 19 kpc from the nucleus, well outside the optical
radius of this galaxy. This filament was independently discovered by
Hameed et al. (2001). Our data show that the [N~II]/H$\alpha$ ratio in
the filament is unlike what is typically seen in H~II
regions. Multi-line imaging slightly shifted in velocity space
suggests that the warm gas is blueshifted by 100 -- 150 km s$^{-1}$
with respect to systemic velocity. A deep long-slit spectrum obtained
with the MSSSO 2.3m telescope confirms these results. The recent study
of NGC~7213 by Hameed et al. (2001) reveals a highly disturbed H~I
system suggesting a past merger event. The line-emitting filament
seems to coincide spatially and kinematically with the main H~I
filamentary structure.

\begin{figure}
\vskip 6in
\caption{ H$\alpha$ + [N~II] $\lambda$6583 + continuum image of
NGC~7213.  A line-emitting filament is detected 19 kpc SW of the
nucleus, well beyond the optical extent of NGC~7213. The existence of
this filament has been independently confirmed by Hameed et
al. (2001).}
\end{figure}
 
A recent TTF study by our group of the radio-quiet quasar MR~2251--178
has revealed a very extended nebula centered on and photoionized by
this quasar (Shopbell, Veilleux, \& Bland-Hawthorn 1999). A
spiral-like complex extending more or less symmetrically over $\sim$
200 kpc is detected in our TTF H$\alpha$ image (Fig. 5; see also
Shopbell, Veilleux, \& Bland-Hawthorn 1999). Narrow-band images
obtained at slightly different wavelengths reveal a large-scale
rotation pattern which is in the opposite sense as that seen in the
inner region of the galaxy (see also Bergeron et al. 1983;
N{\o}rgaard-Nielsen et al. 1986). As discussed in detail in Shopbell
et al., the large and symmetric morphology of the gaseous envelope and
its smooth large-scale rotation suggest that the envelope did not
originate with a cooling flow, a past merger event, or an interaction
with any of the galaxies in the field. Shopbell et al. favor a model
in which the extended ionized nebula resides within a large complex of
H~I gas centered on the quasar.

\begin{figure}
\vskip 6in
\caption{Deep H$\alpha$ image of the field surrounding the quasar
MR~2251--178.  A bright star (S), a nearby cluster galaxy (G1), and a
number of emission-line knots from Macchetto et al. (1990) have been
labeled. North is up and east to the left. The lowest contour in this
figure represents a surface brightness level of $\sim 1 \times
10^{-17}$ erg~s$^{-1}$~cm$^{-2}$~arcsec$^{-2}$. See Shopbell,
Veilleux, \& Bland-Hawthorn (1999) for more detail.}
\end{figure}

\subsection{Summary and Further Improvements}

We have presented some results from our on-going deep emission-line
survey of nearby galaxies with the TTF on the AAT and WHT. Our
preliminary data reveal filamentary complexes extending a few kpc
above and below the disks of normal star-forming galaxies. In active
and starburst galaxies, ionized filaments are sometimes seen extending
out to several tens of kpc. Multi-line imaging of these objects reveals
line ratios which are not H~II region-like.  An early analysis of our
results on normal disk galaxies suggests that the extraplanar gas is
primarily photoionized by the highly diluted and filtered radiation
from OB stars in the disk. In active galaxies, the central engine
appears to be the primary source of ionization of the extended nebula.

These data are providing new constraints on the impact of star
formation and nuclear activity on the host galaxies and their
environment. The implementation of a nod-and-shuffle mode (where the
telescope points alternately on target and on a reference sky position
without reading out) and a broad off-band mode (where the off-band
bandpass can be chosen to be as large as 100 \AA\, therefore reducing
the required exposure time in this band) on the TTF should allow us to
significantly improve the sensitivity of our experiment.

\vskip 0.2in

The survey described in this paper is done in collaboration with
Drs. J. Bland-Hawthorn, G. Cecil, K. Freeman, and P. L. Shopbell and
with University of Maryland graduate students S. T. Miller and
D. S. Rupke.  The author acknowledges partial support of this research
by a Cottrell Scholarship awarded by the Research Corporation,
NASA/LTSA grant NAG 56547, and NSF/CAREER grant AST-9874973.

\end{document}